\documentclass[12pt]{spieman}  
\usepackage{amsmath,amsfonts,amssymb}
\usepackage{graphicx}
\usepackage{setspace}
\usepackage{tocloft}
\usepackage{comment}
\usepackage[utf8]{inputenc}
\usepackage{threeparttable} 
\usepackage{caption}
\usepackage{subcaption}
\usepackage{xcolor} 

\title{Micro-Mirror-Devices (MMDs): A New Family  of MOEMS for the Habitable World Observatory}

\author[a,b,*]{Robberto, M.}
\author[a,b]{Gennaro, M.}
\author[a,b]{Kassin, S. A.}
\author[b]{Smee, S. A.}
\author[c]{Gong. C.}
\author[c]{Huffman, J.}
\author[d]{Ninkov, Z.}
\author[d]{Puchades, I.}
\affil[a]{Space Telescope Science Institute, 3700 San Martin Dr., Baltimore, MD 21218, USA}
\affil[b]{Johns Hopkins University, Bloomberg Center for Physics and Astronomy, 3400 N. Charles Street, Baltimore, MD 21218, USA}
\affil[c]{Texas Christian University, Department of Engineering, 2840 West Bowie Street, Fort Worth, Texas 76109, USA}
\affil[d]{Rochester Institute of Technology, Center for Imaging Science, 54 Lomb Memorial Drive, Rochester, NY 14623-560, USA}

\cftpagenumbersoff{figure}
\cftpagenumbersoff{table} 
\begin{document} 
\maketitle

\begin{abstract}
We present a new program aimed at developing a new generation of micromirror devices specifically tailored for astronomical applications, multi-slit spectroscopy in particular. 
We first overview the general characteristics of  Multi-Object-Spectrographs based on the current Digital Micromirror Devices (DMDs), with particular focus on the newly deployed SAMOS instrument at the 4.1~m SOAR telescope on Cerro Pachon. We illustrate the operational advantages of DMD-based instruments and the technical limitations of the  currently available devices, the DMDs produced by Texas Instruments (TI). We then introduce the baseline and target parameters of the new Micro-Mirror-Devices (MMDs) that we plan to develop with the goal of reaching TRL-5 by mid-2029 as required by the Habitable Worlds Observatory (HWO) timeline. We conclude with a brief illustration of the exciting potential of MMD-based spectrographs for an 8~m class space telescope like HWO.
\end{abstract}

\keywords{Spectrometer, DMD, MMD, HWO}

{\noindent \footnotesize\textbf{*}Corresponding author: \linkable{robberto@stsci.edu} }


\section{Introduction}
\label{sect:intro}  
\begin{figure}[H]
\begin{center}
\begin{tabular}{c}
\includegraphics[height=5.5cm]{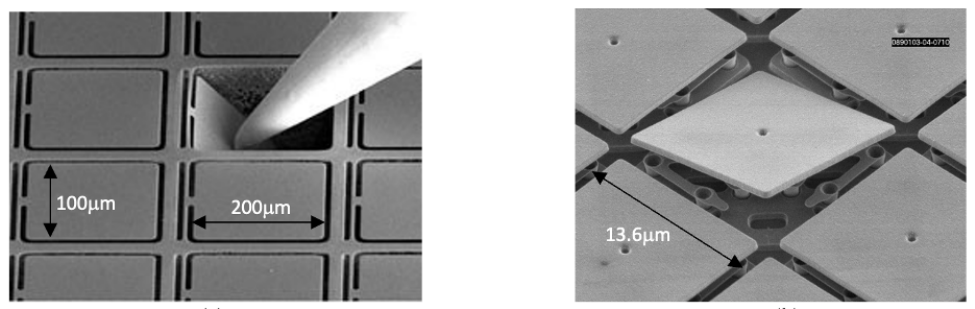}
\end{tabular}
\end{center}
\caption{ 
\label{fig:MOSexample}
Left: zoom-in on a GSFC Microshutter Array, working in transmission
(From \href{https://science.nasa.gov/mission/webb/microshutters/} 
                      {https://science.nasa.gov/mission/webb/microshutters/})
;
right: zoom-in on a Digital Micro Mirror by TI, working in reflection (From \href{https://spectrum.ieee.org/chip-hall-of-fame-texas-instruments-digital-micromirror-device}{https://spectrum.ieee.org/chip-hall-of-fame-texas-instruments-digital-micromirror-device}). Note the different scale of the two images 
} 
\end{figure}

In the last two decades multiplexing techniques based on micro-opto-electro-mechanical systems (MOEMS) have emerged as viable solutions to address the limitations of other conventional techniques for Multi-Object Spectroscopy (MOS) target selection ({ e.g.} punch plates{\cite{2010SPIE.7735E..5AW}}
, {stacked slicing bars\cite{2008SPIE.7018E..0IS}}, fiber positioners{\cite{1998SPIE.3355..828L}}), with the added potential of enabling Integral-Field Unit (IFU)-like capabilities over large areas through Slitlet Stepping
and Hadamard Transform Spectroscopy.
Two types of {MOEMS} devices have been used as rapid reconfigurable slit masks for MOS: microshutter arrays (MSAs) and digital micromirror devices (DMDs). Both families of devices consist of arrays of individually addressable elements that act as optical switches, individually programmed to an "open" or "{closed}" status, to selectively capture the spectroscopic target sources. While MSAs work in transmission, DMDs work in reflection (Fig.~\ref{fig:MOSexample}). 

\section{Historic overview}
\subsection{Early efforts}
Around the year 1999 NASA started the development of { MOEMS} devices as slit selectors for the near-infrared spectrograph {{NIRSpec}} of NGST (later renamed the James Webb Space Telescope). Three development programs were initially funded. Two groups, one at Sandia National Laboratories and another at NASA/Goddard Space Flight Center (GSFC), independently started the development of micromirror devices, whereas a third group, also at GSFC, envisioned and developed the MSAs. At the end of 2001 the MSAs were downselected { largely based on the expectation that they could more easily achieve the 1:2,000 contrast requirement set for NIRSpec \cite{kutyrev_microshutter_2008}}. NASA therefore stopped the micromirror programs. 
The MSAs that eventually flew on JWST are arrays of 171x365 rectangular microshutters 100 by 200~$\mu$m each.  The NIRSpec spectrograph, operating at a temperature of approximately 40~K, uses four MSA quadrants to cover a focal plane area of about 10.5 cm$^2$ in the spectral range between 1 and 5.5~$\mu$m. 
Each individual shutter covers a 0.27'' x 0.53'' field with a 64\% filling factor. { A Next Generation Microshutter Array is now being developed to overcome the limitation of the original microshutters \cite{10.1117/12.3020479}, with prototypes flown on the FORTIS sounding rocket missions \cite{9136702}}. 

\begin{figure}[H]
\begin{center}
\begin{tabular}{c}
\includegraphics[height=5.5cm]{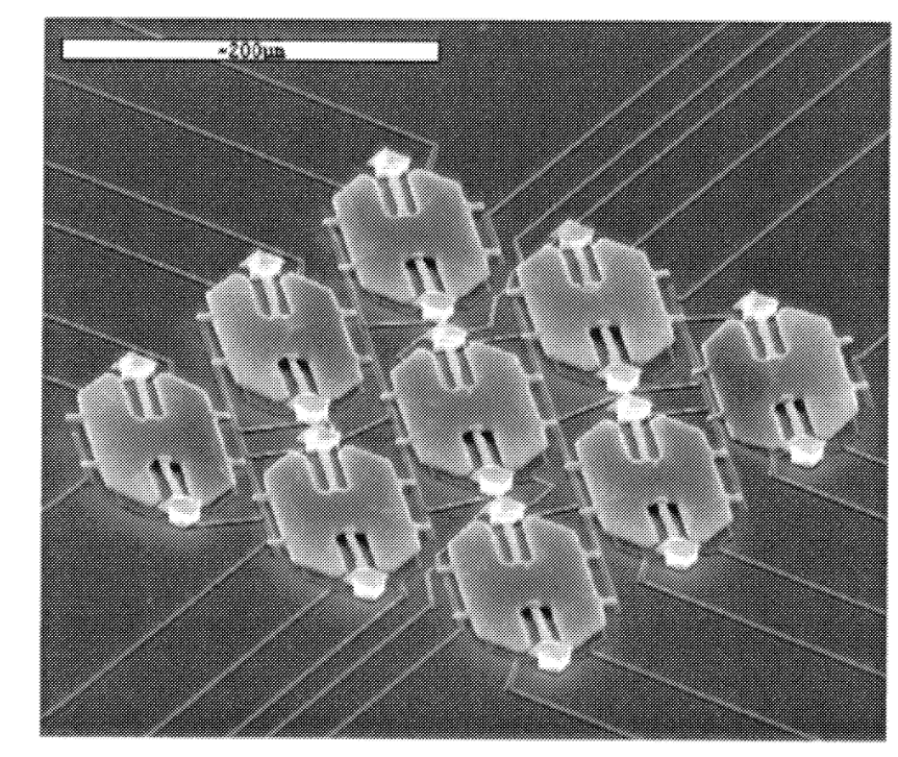}
\end{tabular}
\end{center}
\caption 
{ \label{fig:Dutta}
Scanning Electron Microscope image showing the underlying structure of a 3$\times$3 DMD array developed by GSFC around the year 2000 (from Ref.~\citen{dutta_development_2000}).  Note that TI soon after abandoned this type of layout. } 
\end{figure}

Of the two micromirror teams, the one at GSFC leveraged their concept on the Digital Micromirror Devices (DMDs) developed in the 1980s at TI for projection display applications (Fig.\ref{fig:Dutta}). In the year 1997 TI's DMDs had just reached a level of maturity adequate for introduction to the consumer market. 
Adapting the TI design to the requirements of JWST, the main challenges were to expand the pixel size from $17\,\mu$m $\times17~\mu$m to $100~\mu$m$\times 100~\mu$m while ensuring reliability in cryogenic conditions and optical quality in terms of both mirror flatness and contrast \cite{dutta_development_2000}. 

\subsection{DMD evolution}
In the last 25 years TI has continued to develop DMD technology, launching families of devices with increasingly outstanding  performance.
A modern DMD consists of an array of up to 8.8 million square micromirrors placed on top of a CMOS circuitry which allows the mirrors to be individually addressed and tilted into one of two stable states. During operation, a typical micromirror is tilted at + and - 12 degrees from the device normal at frequencies of several kilohertz. The rapid binary modulation between the two states is used to create hundreds of gray-scale digital pictures per second, i.e. a series of “photograms” that eventually create a color video when cyclically illuminated by a synchronized RGB light source. 

DMDs are a mature technology with impressive reliability and are generally regarded as the most successful {MOEMS} devices ever developed.  Already in the year 2016 TI had delivered 40 million DMDs around the world \cite{TI_2016}.  For a projector operating for 10 years,  each micromirror  goes through approximately $10^{12}$ cycles. If the DMD has 1 million micromirrors, then the total number of cycles for all mirrors is about $10^{18}$ (a quintillion!) cycles. The impact of DMDs on the film industry has been so significant that in 2015 the inventor Larry Hornbeck was awarded with an Oscar ``for his contribution to revolutionizing how motion pictures are created, distributed and viewed''. 
Even if projection applications set requirements rather different from those of astronomical instruments, TI DMDs have been considered and adopted by astronomers, in particular in Multi-Object Spectrographs.

\section{DMD-based Multi-Object-Spectrographs}
\subsection{Design strategy}
\begin{figure}[H]
\begin{center}
\begin{tabular}{c}
\includegraphics[height=5.5cm]{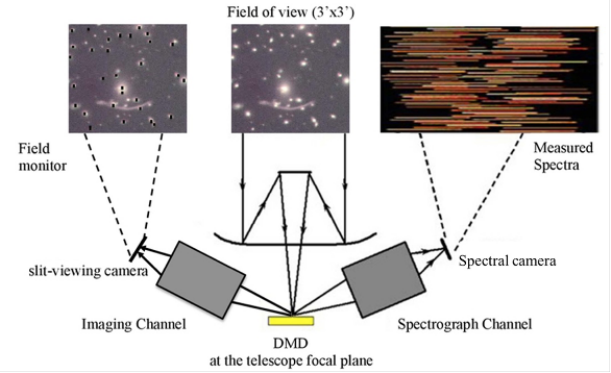}
\end{tabular}
\end{center}
\caption 
{ \label{fig:MOS_classic}
Schematic illustration of a typical layout for a DMD-based MOS.} 
\end{figure}

A typical design \cite{2009SPIE.7210E..0AR} for a DMD-based MOS utilizes both the positive and negative tilt states of the micromirrors as shown in Fig.~\ref{fig:MOS_classic}.  The telescope focal plane is projected on the DMD, each micromirror representing a potential slit reflecting light from selected target sources towards the spectrograph channel while the remaining micromirrors redirect the field-of-view towards an imaging channel. The imaging channel thus functions as a slit-viewing device for acquisition and monitoring, but can also perform deep imaging in parallel with the spectroscopic integration. The locations and sizes of slits are fully determined by the pattern deployed on the DMD. 

Several DMD-based MOS have been built and operated at ground-based telescopes including RITMOS\cite{meyer_ritmos_2004} in the visible {at the 24'' Mees telescope}, IRMOS\cite{2004SPIE.5492.1105M} in the near-IR { at the KPNO 2.2~m and 4-m Mayall telescopes}, and recently SAMOS\cite{samos_2016} also in the visible {at the 4.-m SOAR telescope}. 
SAMOS uses a CINEMA  DMD, format {$1080$}$\times 2048$ with 
{ 13.0~$\mu$m  mirror size and 13.68~$\mu$m center-to-center spacing, corresponding to a geometric filling factor of 90.3\%}. This is the same device envisioned for the SPACE\cite{2009ExA....23...39C} mission proposal that was selected by ESA and later morphed in the NISP instrument of Euclid, as well as for the GMOX feasibility study \cite{2016SPIE.9908E..52B,2016SPIE.9908E..4ZS,2016SPIE.9908E..49G} funded by the Gemini Observatory for a next generation facility instrument. 

\subsection{Current state-of-the-art: SAMOS}
\begin{figure}[H]
\begin{center}
\begin{tabular}{c}
\includegraphics[trim={6cm 3.cm 5cm 0},height=7.5cm]{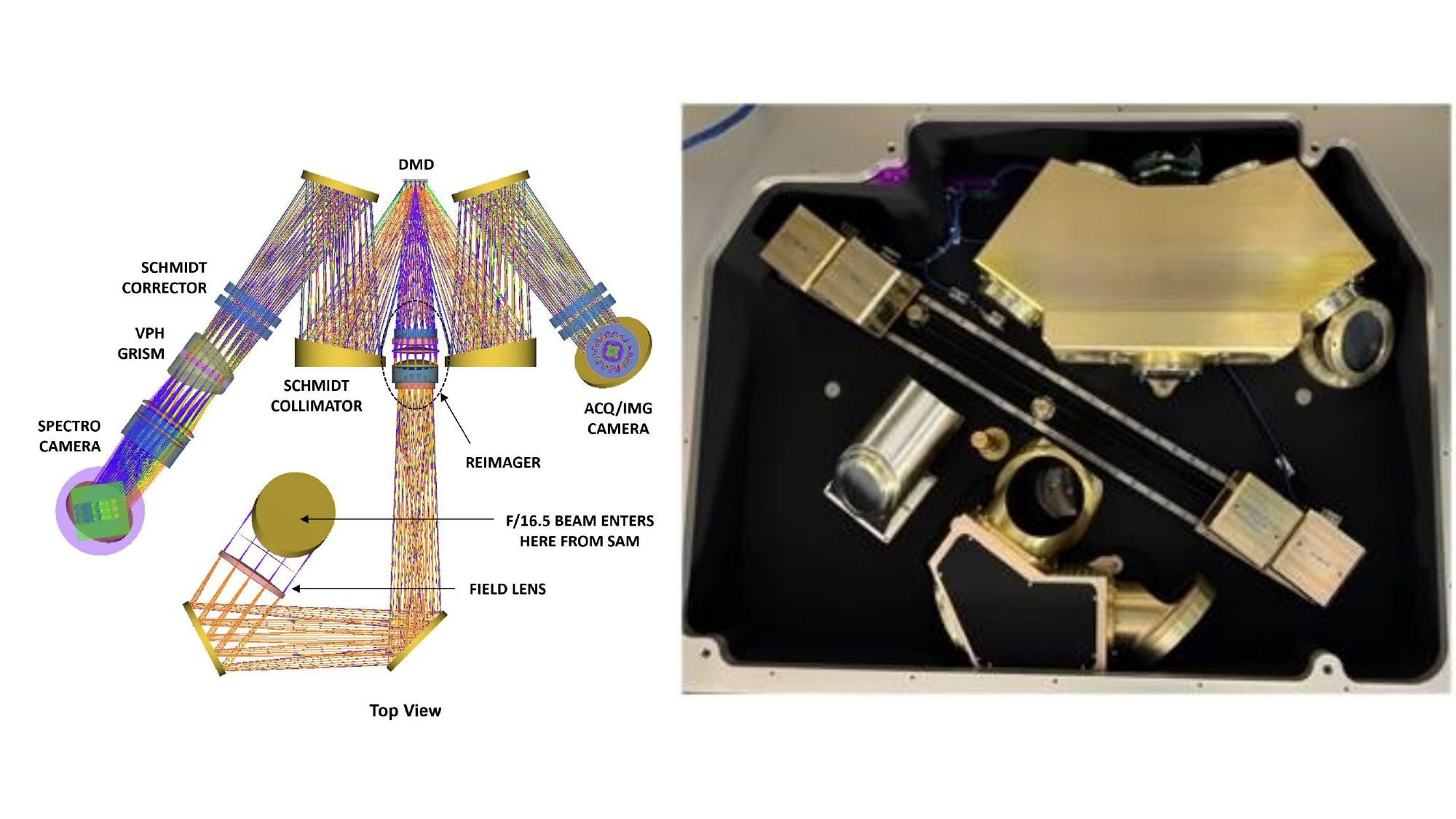}
\end{tabular}
\end{center}
\caption 
{ \label{fig:SAMOS_layout}
Left: top-view of the SAMOS optical design\cite{2016SPIE.9908E..4ZS}; the layout is compact and highly folded to fit the reduced envelop available at the SOAR Adaptive Optics Module SAM focal  plane. Right: corresponding top-view of the as-built instrument.}
\end{figure}

SAMOS (Fig.\ref{fig:SAMOS_layout}) has been recently deployed at the 4.1 m SOAR telescope on Cerro Pachon, Chile to fully exploit the enhanced seeing quality delivered by the Ground Layer Adaptive Optics system (SAM) over a $3\times3$ arcmin field of view\cite{tokovinin_soar_2016}. SAMOS { optics reimage this field of view over the central $1080\times1080$ mirrors, covering the 400 - 950 nm visible range} at $R\sim3000$ for 0.36" slits with two exposures in the low resolution VPH { transmission} gratings. Two other gratings reach $R \simeq 10,000$ in the spectral regions around the blue H$\beta$-[OIII] and red H$\alpha$-[SII] lines, respectively. The spectra are collected { with a 1.125 mirror/pixel sampling} by the SAMI camera  normally dedicated to direct GLAO imaging, based on the 4096$\times$4112 CCD231-84, while the SAMOS imaging channel {relays the $3'\times3'$ field to}  a high-quality 1K$\times$1K e2v 47-10 CCD. The imaging channel is equipped with two filters wheels hosting a $griz$ filter set and four filters centered on the major H$\alpha$, H$\beta$, O[IIII] and S[II] lines. 

\section{The exceptional capabilities of DMD-based spectrographs}
The operations of an advanced DMD-based instrument like SAMOS showcase the exceptional versatility of DMD-based spectrographs.

\subsection{Target Acquistion}
For target acquisition, the targets can be selected in real time at the telescope through "point and click" on the acquisition image, or more appropriately can be determined in advance. In this second case, the astronomers prepare Astropy region files listing the right ascension and declination of each target, together with the width, length and orientation of each corresponding slit. At the telescope, a short acquisition image is processed in real time to extract the point sources and cross-match their positions with the Gaia catalog. This immediately provides the astrometric WCS solution of the image, allowing the original region file to be converted from celestial coordinates to CCD pixel units. 
\begin{figure}
\begin{center}
\begin{tabular}{c}
\includegraphics[height=7.5cm]{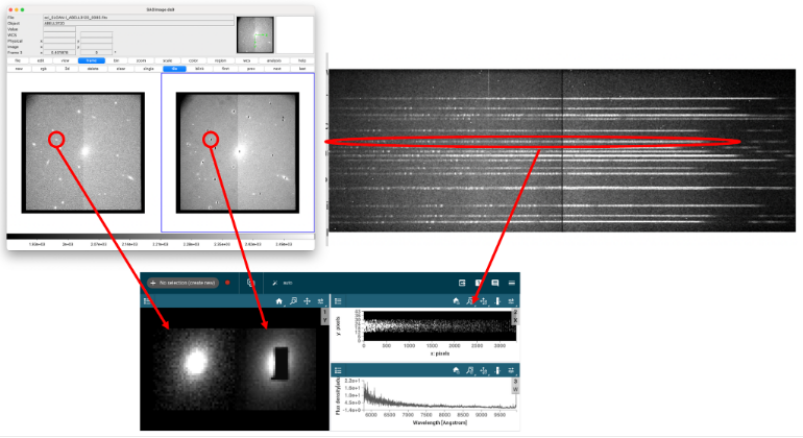}
\end{tabular}
\end{center}
\caption 
{ \label{fig:SAMOS_acquistion}
SAMOS test data taken during the commissioning run in October 2024 for the galaxy cluster Abel 3120: top row, from left to right: the first target acquisition image on the imaging channel; the same field with the slits configured based on the pre-selected RADEC coordinates of the targets; the simultaneous MOS spectra, raw data, 900s integration. Bottom row: the same data for a random galaxy displayed after quick-look reduction using the JWST MOSVIZ Data Analysis Tool. Full moon conditions, no GLAO laser guide.}

\end{figure} 
The conversion from pixels to micromirrors is done using a calibration map previously determined from a regular grid of illuminated micromirrors. The micromirrors corresponding to the targets, with their corresponding slits, are therefore identified and redirected to the spectroscopic channel. The full target acquisition sequence, once the initial image has been taken, requires a few seconds, in practice a negligible overhead in comparison to the guide-star and laser-guide acquisition time. 

\subsection{Science Observations}

\begin{figure}
\begin{center}
\begin{tabular}{c}
\includegraphics[trim = {7cm 4cm 5cm 5cm},height=5.5cm]{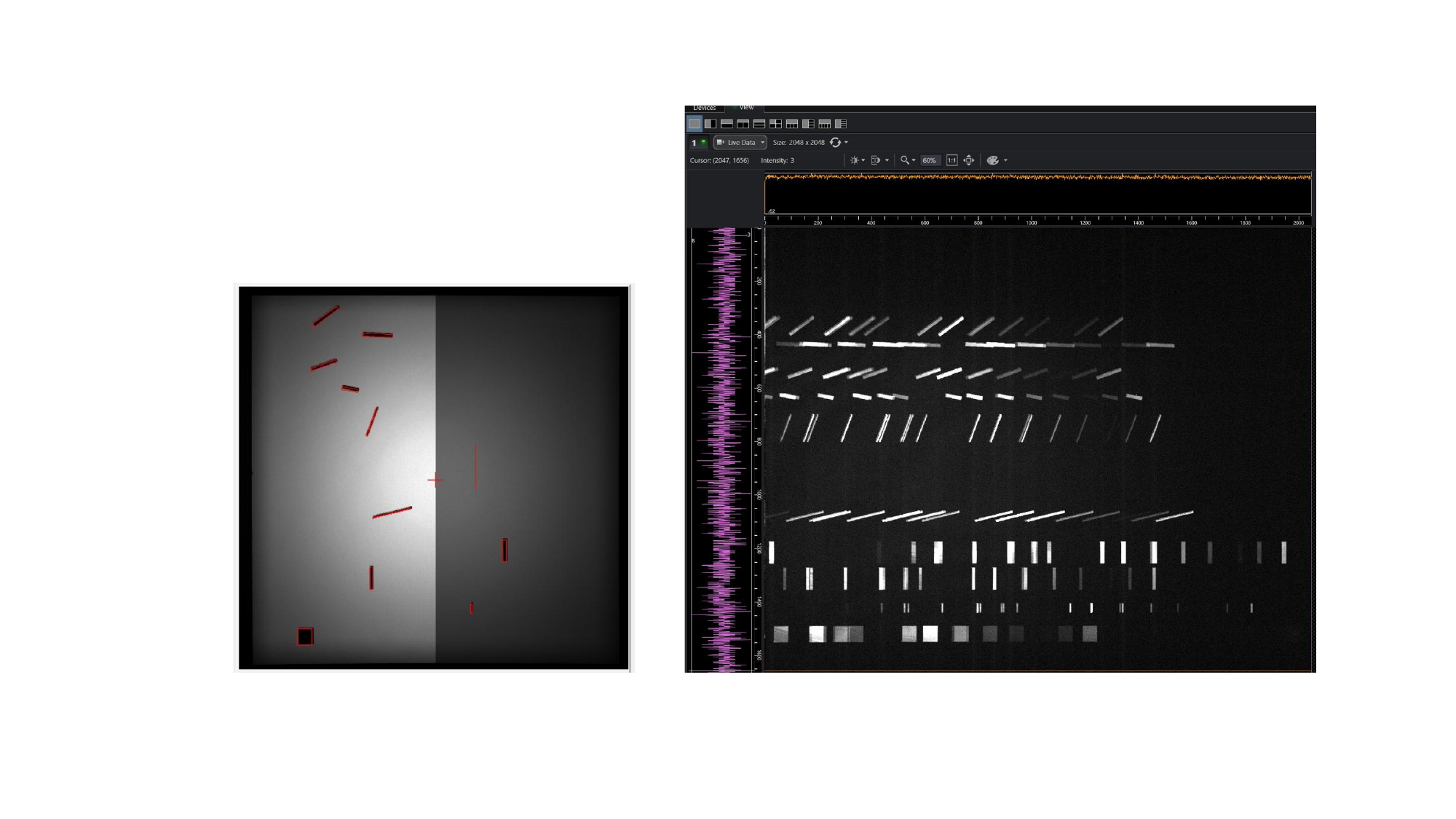}
\end{tabular}
\end{center}
\caption 
{ \label{fig:SAMOS_Crazy}
Left: SAMOS lab-test images demonstrating the versatility of the DMD, capable of generating slits with virtually any shape in less than 3 seconds. Left: { randomly generated slits seen by the imaging channel, 1K$\times$ 1K CCD}. Right: { same slit pattern seen by the spectroscopic channel, with the spectrum produced by an arc calibration lamp on a test 2K$\times $2K CCD}.}
\end{figure}

The spectroscopic and imaging exposures can start and proceed in parallel, with the possibility of exchanging gratings and filters asynchronously (Fig.\ref{fig:SAMOS_acquistion}).
Deep imaging of the field can thus be obtained simultaneously to the spectroscopic acquisition, with only the relatively bright spectroscopic targets masked out.
The slit width and therefore the spectral resolution can be adjusted in real time, globally or for specific targets, to match the actual seeing conditions and the performance of the SOAR Ground Layer Adaptive Optics Correction. 
It is possible to use replicated patterns optimized to the flux of the targets, swapping intermediate brightness targets while insisting on the faint ones, or switch sources if their traces overlap. With a DMD pattern load taking about 3 seconds, the main factor limiting observing efficiency is in general the detector readout time. 
The full toolset of multi-slit spectroscopy, including refined observing modes like e.g. nod-and-shuffle for optimal sky cancellation, can also be readily implemented with limitless versatility (Fig.~\ref{fig:SAMOS_Crazy}).

\subsection{Calibration}
The comparison of images taken without and with the slits allows one to precisely determine the fraction of light going through each slit; moreover, the availability of the standard $griz$ filters allows one to cross-calibrate in real-time all targets against e.g. the PanStarrs or SkyMapper catalogs for immediate and accurate spectrophotometry. Removing the grating from the spectrograph light path and inverting the DMD pattern allows one to visualize the image also on the spectrograph, facilitating the cross-matching of the targets vs. their corresponding slits and traces.

Each slit pattern in DMD units, generated for an actual telescope pointing (WCS solution), is saved to be reloaded at the end of the night to calibrate flat-field and dispersion. If a very dense slit pattern is desired to maximize the number of targets, the small pincushion curvature of the slit traces (about 1/1000 over the length of the spectra) can complicate the extraction of the individual traces. In this case, one can simply split the DMD slit pattern in two sets to generate ``even'' and  ``odd'' traces, taking two separated flats each one having enough space between traces to extract them without any contamination.

\subsection{Multi-IFU-like modes}
\begin{figure}[H]
\begin{center}
\begin{tabular}{c}
\includegraphics[height=10.5cm]{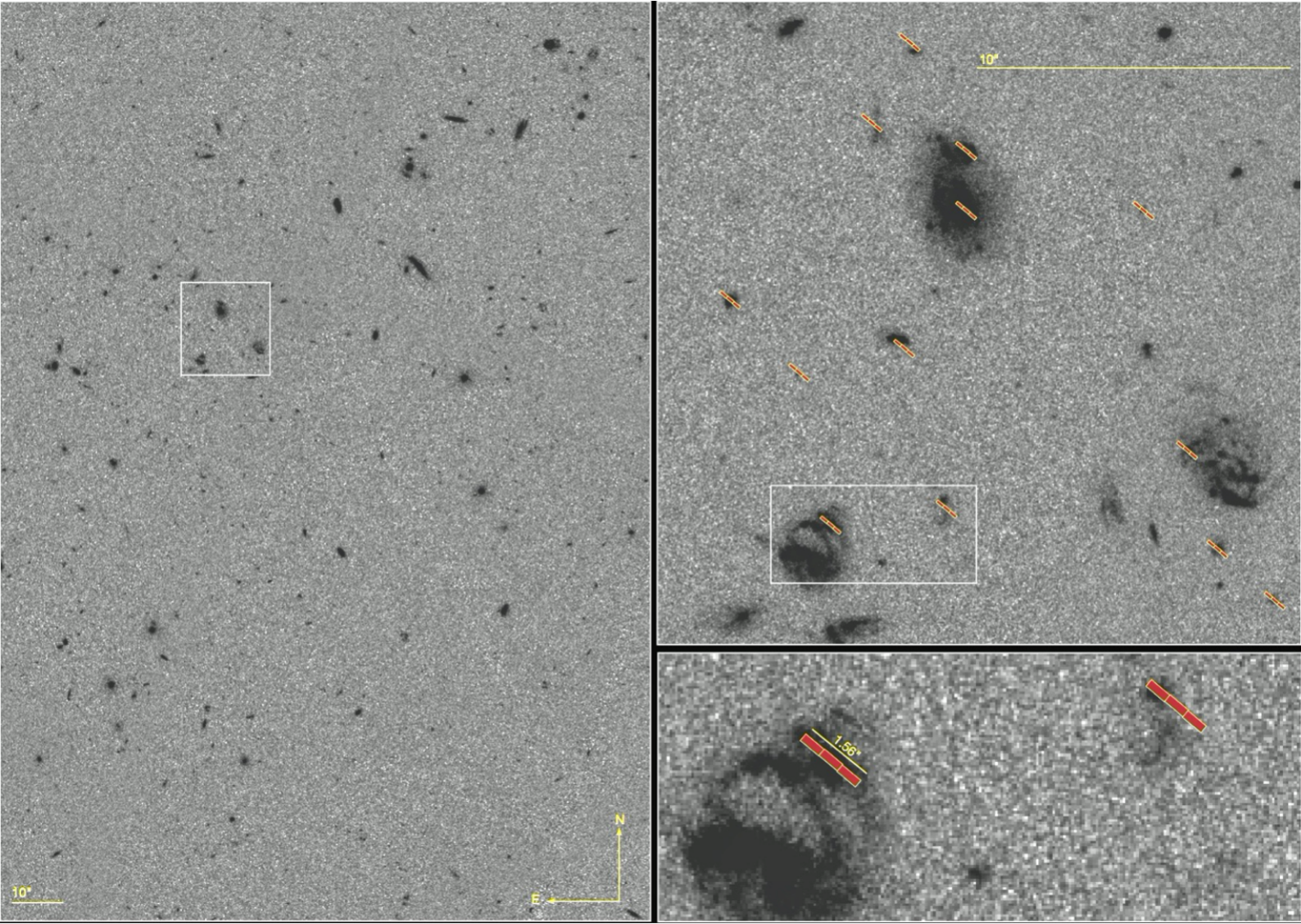}
\end{tabular}
\end{center}
\caption 
{ \label{fig:SlitletScan}
A zoomed-in view of several 3-shutter slitlets configured open on MSA science sources of interest in the Hubble Ultra Deep Field (HST ACS F606W image). From 
\href{https://jwst-docs.stsci.edu/jwst-near-infrared-spectrograph/nirspec-observing-modes/nirspec-multi-object-spectroscopy\#gsc.tab=0}
{https://jwst-docs.stsci.edu/jwst-near-infrared-spectrograph/nirspec-observing-modes/nirspec-multi-object-spectroscopy{\#}gsc.tab=0}
}
\end{figure}

It is possible to adapt the DMD pattern for Integral Field Observations.
Slitlets stepping \cite{mackenty_multi-object_2000, 2025AAS...24544506P,Kassin_2025} have been recently implemented on JWST/NIRSpec\cite{NIRSPEC}
 to obtain full datacube reconstruction of a subfield within the MSA area. Scanning a slitlet (Fig.~\ref{fig:SlitletScan}) to cover e.g. the extension of a compact galaxy provides the same information of a full IFU, with a penalty given by the time needed to perform the scan. In the case of NIRSpec, a $3''\times3''$ IFU area can be covered scanning a $1\times6$ MSA slitlet 11 times across the target. A NIRSpec scan may need to be repeated with small dithers to recover spatial resolution, but this is compensated by the factor of $\approx 4$ in throughput between the NIRSpec MSA vs. IFU. All considered, depending on the number of individual slits and number of targets in the field, there is a net efficiency gain using slitless scan as ``multi-headed'' IFU vs. the image-slicer IFU that makes this mode optimal for observing e.g. a cluster of galaxies. Observing efficiency and signal-to-noise considerations may dictate if slitlet stepping is better done by shifting the pointing of the telescope or the slit pattern, but the instrinsic instrument versatility allows for hybrid modes, where e.g. a number of relatively extended targets is observed in slitless-stepping mode while others, fainter and more compact, are observed in regular slit mode during the same observation sequence\cite{mackenty_multi-object_2000}. 

\begin{figure}[H]
\begin{center}
\begin{tabular}{c}
\includegraphics[height=7.5cm]{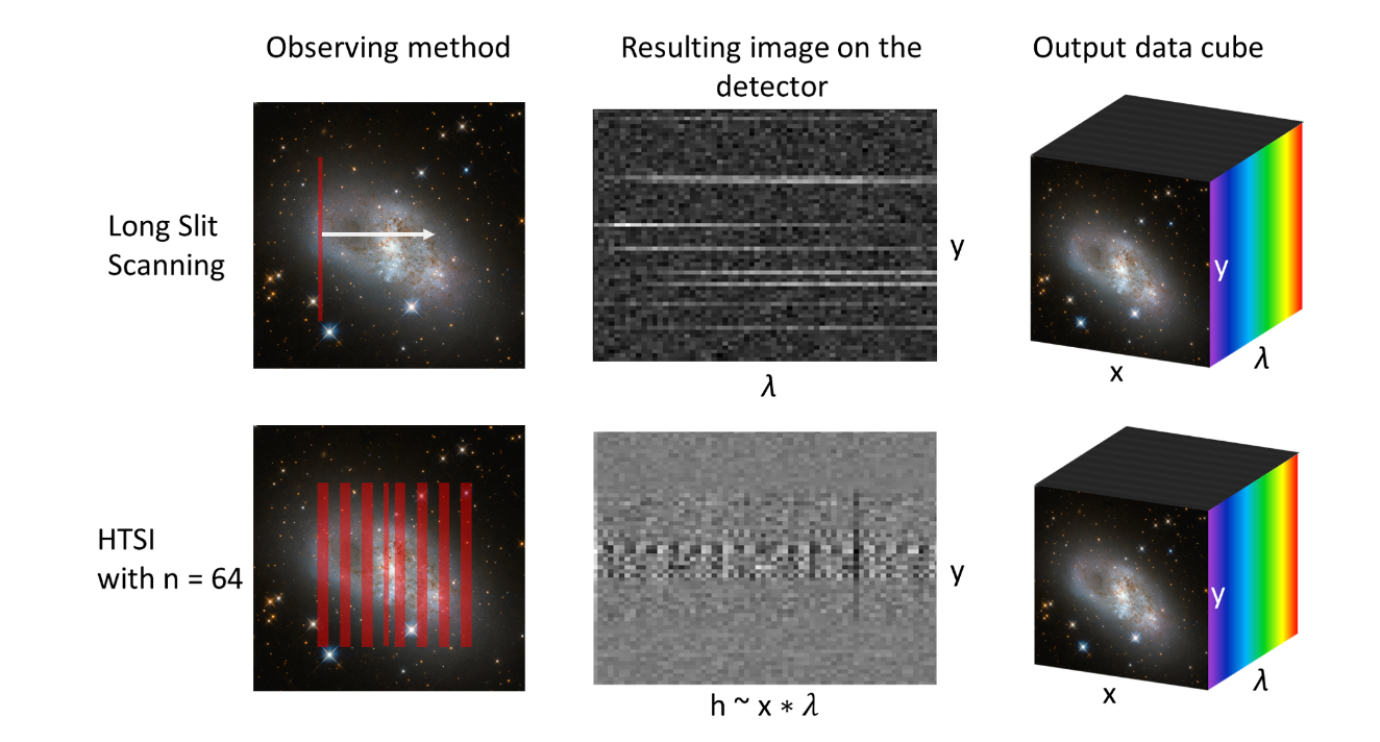}
\end{tabular}
\end{center}
\caption 
{ \label{fig:OramFig2}
A comparison of long slit scanning and Hadamard Transform Spectral Imaging (HTSI) methods (from ref. \citen{oram_modeling_2022}) 
}

\end{figure}

A further gain in Signal-to-Noise can be obtained if the slitless scan is performed by opening multiple slits at the same time according to patterns implementing the coding scheme of Hadamard matrices (Fig.~\ref{fig:OramFig2}), a technique demonstrated with the IRMOS DMD-spectrograph at Kitt Peak\cite{fixsen_spectroscopy_2009}. For each given ``barcode'' pattern, the signal seen by a pixel is the superposition of $n$ unknown contributions (spaxels) coming from different slits, at different wavelengths. Note that for each pattern, about 50\% of the spaxels are always masked out. The ensamble of $n$ unknown can be determined if $n$ different superpositions/patterns are generated, resulting in a matrix of $n$ linear combinations of $n$ unknowns that can be directly inverted. The gain in Signal to Noise (Fellget advantage) depends on $n$, on the type of matrix and on the relative ratio of read vs. shot noise, the read-noise limited case providing the largest advantage over simple slit scan\cite{nitzsche_noise_2003, streeter_optical_2009,Nijim_HTS_2025}. 

\section{Beyond TI DMDs: the Micro-Mirror-Devices (MMDs)}

Extensive characterization campaigns have demonstrated that DMDs (both type Cinema 2K \cite{2010SPIE.7731E..30Z}  and DLP7000\cite{travinskya_evaluation_2017}) are extremely robust to shock, vibrations, cryogenic temperatures and harsh radiation environment, and can therefore be assessed at the TRL 5 required for consideration in a space-based instrument. Still, even though a DMD-based MOS presents significant advantages over other multiplexing approaches, the current DMD design presents features that limit their potential for a future space mission like HWO.

TI DMDs are generally produced for digital imaging and their main characteristics reflect this particular type of application. They are designed for spatial light modulation, flipping each individual mirror at 20-30~KHz, and through the years the tendency has been to shrink the mirrors to smaller and smaller sizes as this increases the resolution for a given chip size, facilitates the highest frequency operations, and through miniaturization opens new consumer market possibilities (e.g. pico-projectors). Unfortunately this direction is not optimal for astronomical applications requiring larger physical format, hence our consistent interest in the CINEMA  devices, admittedly not the latest generations of DMDs but still the largest model with the largest micromirror size. With TI being historically resistant to customizing  DMDs for a niche application as astronomy, designing  a spectrograph based on TI DMDs poses a number of constraints, listed in Appendix~\ref{sec:Appendix}. 

A new micromirror design optimized for astronomical applications can overcome all the limitations of commercial DMDs enabling an entirely new class of high-performance instruments. We have envisioned such a new generation of devices, dubbed Micro-Mirror-Devices (MMDs), and started the process to seek the needed funding to implement a development plan aimed at delivering by mid-2029 our baseline devices at TRL-5. Our team members Gong and Huffman had been instrumental in developing the CINEMA 2K DMD devices, so our effort leverages decades of experience matured  at TI. In comparison to DMDs, our proposed MMDs  will be relatively straightforward to fabricate due primarily to the larger format and low operational frequency. Still, the operational requirement in terms of size, reliability, and optimization for a broad wavelength range, UV in particular, poses challenges that require a phased R\&D approach, to be presented in a future paper. In order to enable MOS with unprecedented capabilities, our target micromirror devices  are expected to meet the specifications listed in Table~\ref{tab:MMDs} below {, with the current parameters of the TI CINEMA  DMD shown for comparison}.

\begin{table}[ht]
\caption{MMDs baseline and goal parameters} 
\label{tab:MMDs}
\begin{center}       
\begin{tabular}{|l|c|c|c|} 
\hline
\rule[-1ex]{0pt}{3.5ex}  Parameter & MMD-baseline & MMD-goal & {TI CINEMA } \\
\hline\hline
\rule[-1ex]{0pt}{3.5ex}  Pitch&	$30~\mu$m &	$100~\mu$m  &	{$13.68~\mu$m}  \\
\hline
\rule[-1ex]{0pt}{3.5ex}  Fill Factor & $>91\%$ & $>96\%$ &	{90.3\%}\\ 
\hline
\rule[-1ex]{0pt}{3.5ex}  Format&	1K$\times$ 1K & 2K$\times$ 2K (buttable 2$\times$2) &	{$1080\times2048$ }\\
\hline
\rule[-1ex]{0pt}{3.5ex}  Range &	NUV-VIS &	FUV-VIS-NIR  &	{VIS} \\
\hline
\rule[-1ex]{0pt}{3.5ex}  Tilt angle	&12$^\circ$ to 15$^\circ$ & 15$^\circ$ &{12$^\circ$}\\
\hline

\rule[-1ex]{0pt}{3.5ex}  Mirror reflectivity&	96\%	&96\% &	{96\%}\\
\hline
Contrast Ratio       & 1:10,000 (blue) &1:20,000 (blue)&{1:5,840 (blue)}\\
                     &  1:5,000 (red)   &1:10,000 (red)  &{1:2,790 (red)\cite{piotrowski_-situ_2024}} \\
\hline
\rule[-1ex]{0pt}{3.5ex}  Reconfiguration time	& $< 4$ s	& $< 6$ s &{$\sim 3$ s}\\
\hline 
\end{tabular}
\end{center}
\end{table} 
Table~2 summarizes the basic characteristics of a MOS spectrograph assuming an 8~m diameter telescope { with a reimaging optics} feeding either our baseline or goal MMDs with an f/4 optical system, {i.e. the same focal ratio of SAMOS}. The extremely large number of mirrors, up to 16 million of for a 4K$\times 4$K mosaic, allows for enormous efficiency gains. The multiplexing advantage achieved by two parallel MOS spectrographs built around 4K$\times 4$K MMD mosaics is comparable to that achieved by PFS on Subaru (2400 fibers, 1.05" diameter). At the same time, the same MOS can enable observing strategies { which are} highly competitive vs. state-of-the-art multi-IFU instruments.

\begin{table}[ht]        
\begin{center} 
    \begin{threeparttable}[b]
    \label{tab:MOS}
        \caption{Basic parameters of a MMD-based MOS on a 8m class telescope (HWO)} 
      
        \begin{tabular}{|l|c|c|c|} 
            \hline
            \rule[-1ex]{0pt}{3.5ex}  MMD Pitch & $30 \mu$m  & $100~\mu$m  & $100~\mu$m  \\
            \rule[-1ex]{0pt}{3.5ex}  MMD Format&	1K$\times$ 1K & 2K$\times$ 2K  & 4K$\times$ 4K (butted 2$\times$2)  \\
            \hline
            \rule[-1ex]{0pt}{3.5ex}  Scale &	0.19''/mirror &	0.64"/mirror & 0.64"/mirror  \\
            \hline
            \rule[-1ex]{0pt}{3.5ex}  FoV	&$3.3'\times3.3'$ & $22'\times22'$ & $44'\times44'$ \\
            \hline
            \rule[-1ex]{0pt}{3.5ex}  Nr of parallel MOS slits\tnote{$^1$}	&104 & 688 &1375 \\
            \hline
            \rule[-1ex]{0pt}{3.5ex}  Nr of parallel IFU-like areas\tnote{$^2$}	&51 & 338 &676 \\
            \hline
            \end{tabular}
            \begin{tablenotes}
               \item [1] Assumes $2"$ long slitlets per target.
               \item [2] Assumes $3.9"\times 3.9"$ IFU-like target area.
            \end{tablenotes}
    \end{threeparttable}
 \end{center}   

\end{table} 

\section{Conclusion}
\label{sec:Conclusion}

As demonstrated by the recently deployed SAMOS spectrograph at the 4.1 m SOAR telescope, the Digital Micromirror Devices produced by TI enable Multi-Object-Spectrographs with unique capabilities, especially in terms of operational efficiency and versatility.
Still, DMDs are optimized for high speed light modulation and this ultimately constrains the instrument design of a MOS spectrograph to a relatively narrow set of parameters.

In this report we have introduced our new program aimed at developing a new generation of Micro-Mirror-Devices (MMDs) specifically tailored for astronomical applications, multi-slit spectroscopy in particular. Overcoming the limitations of TI DMDs, we plan to raise these devices to TRL-5 by the year 2029 to be considered for the HWO. MMDs will expand by orders of magnitude the scientific capabilities (field coverage, multiplexing, wavelength coverage...) of TI DMDs, enabling transformative science at a fraction of the complexity and costs of other technologies.
%
%

\appendix
\section{Considerations on TI DMDs for MOS design}
\label{sec:Appendix}
In this section we briefly summarize the main factors that need to be considered designing a DMD-based astronomical instrument. Our MMD development program aims at overcoming the limitations presented here.
\begin{enumerate}
\item{Mirror size\\}
The largest mirror size of TI DMDs is $13~\mu$m, with $13.68~\mu$m center-to-center pitch.  By comparison, the physical size of the individual MSA slits of JWST/NIRSpec is $100\times200~\mu$m.  A small mirror size tends to oversample the Point Spread Function and therefore even slits with the smallest angular size on the sky often require more than one mirrors across. Ideally, one could illuminate the DMD with a fast f/\# to shrink the PSF size, with the added advantage of increasing the field of view. However, 
for perpendicular DMD illumination, the 12$^\circ$ mirror tilt sets a lower limit to the f/\# at around f/2.5, in order to prevent overlapping the incoming and reflected beams. SAMOS illuminates perpendicularly the CINEMA 2K DMD at f/4, about the practical limit, that with a 4.1~m aperture telescope corresponds to about 0.18''/micromirror. At an 8m telescope this would correspond to 0.09"/micromirror. An alternative approach is to illuminate the DMD { with an angle larger than 24$^\circ$ to increase the separation of the reflected beam}, but this significantly increases the complexity of  the optical design \cite{content_atlas_2018}.  


\item{Format\\}
There are limited format options in the TI { product line}.  The rectangular shape adopted by video industry standards is not always optimal for MOS applications, adding constraints to the optical design. As an example, SAMOS adopts a $2048\times 1080$ DMD but exploits only the central 1K$\times$1K since the SOAR Adaptive Optics module (SAM) delivers a square field of view. 

\item{Tilt angle\\}
The tilt angle of CINEMA  DMDs is 12$^\circ$, and the same value is used for other types of DMDs with  10.8$~\mu$m and 7.6/7.56$~\mu$m micromirror pixel sizes. It increases, however, to 14.5$^\circ$ for DMDs with the 9$~\mu$m and to 17$^\circ$ for models with 5.4$~\mu$m pixel size. Larger tilt angles allow for increased separation of the beams and faster optics, improving both contrast and areal coverage. DMDs with larger tilt angles also showcase higher fill factor e.g. from 92\% to 97\% moving from DMDs with 12$^\circ$ to 14.5$^\circ$ tilt angles. 
While increasing the tilt angle maintaining full reliability requires profound understanding of the mechanical properties of the alloy materials, creating torsion hinges with flexure angles of $10^\circ-15^\circ$ is generally less demanding than reaching the  $90^\circ$ angles needed by MSAs.

\item{Mirror Reflectivity\\}
TI utilizes a proprietary Al coating with various metal doping to ensure long life-time of the mirrors oscillating at high speeds. The reflectivity is ~89\% in the visible (photopic) range, 420-680 nm. For astronomical applications, one would like to optimize the mirror coatings, enhancing performance especially in the UV and NIR.  Vorobiev et al. (2018) have shown that existing TI DMDs can be re-coated to improve reflectivity without damage, but the implications on device lifetime are unclear. A new generation of micromirror devices designed for astronomy should adopt coatings better matched to selected wavelength ranges. For the FUV, one could envision protecting  high-purity Al with a thin film of LiF\cite{Quijada_advanced_2022} or AlF$_3$\cite{quijada_environmental_2023}.  

\item{Surface optical quality\\}
In principle, the semiconductor processing required to build the mirror structure, in particular the thin surface layer oscillating at very high frequency, can result in the mirrors deviating from a perfectly flat plane. TI has developed a { process} designed to minimize the non-flatness or non-planarity of the mirror and today non-flatness does not seem to represent an issue for TI DMDs. {Larger format mirrors are in principle more sensitive to flatness errors, but for very-low operation frequency applications one may adopt thicker and light-weighted 3-d mirror support structures to preserve pristine optical quality}. The team at NASA/GSFC worked on this issue presenting different planarization processes suitable for large format micromirrors\cite{zheng_planarization_2002}. 

\item{Window Transmission\\}		 	 	 		
The TI DMD micromirrors need to operate in an inert environment for long durability. DMDs are therefore encased in a Kovar container with a hermetically sealed, CTE matched window. TI uses two types of borosilicate glass, either Corning 7056 or  Corning Eagle XG. A/R Coating is typically optimized for the 400-700nm visible range of the human eye, with typical transmittance (double pass) about 96\%.  There is clearly room for improvement with more advanced multilayer coatings. Also, borosilicate is opaque to UV radiation and therefore different window materials are needed. While window replacement is doable, for space applications one could consider operating without a window either by packaging the DMD with a (re)movable window, or making the DMD part of a larger, sealed opto-mechanic subsystem including parts of the optical elements.

\item{Contrast\\}
Contrast, i.e. the fraction of background light leaking through the DMD when all elements are tilted to “off” state, has received a lot of attention in the past. Contrast depends on a variety of factors including wavelength, f/\# of the incoming beam, location and size of the entrance and exit pupils, control of reflections and stray light\cite{piotrowski_optical_2023}. An {often} neglected factor is that contrast is also a strong function of the quality of the device, since faulty open slits leak light that once dispersed contaminates all pixels falling in their spectral trace. { As an example, according to the STScI JDOX documentation for the JWST/NIRSpec spectrograph, the MSA flux leakage can accumulate to a level that is closer to one part in $\sim 50$, on average\cite{2016jdox.rept......}}.
End-to-end contrast measures are the most valuable, and for the CINEMA  DMD we have recently published “in-situ” results with SAMOS\cite{piotrowski_-situ_2024}. We find the median contrast with the low resolution blue grism (400-600 nm) to be 5840:1 while the median contrast with the low resolution red grism (600-950 nm) is 2790:1, surpassing the original requirements for JWST (2000:1).
The two main sources of contrast losses in TI DMDs, scattering and diffraction, are discussed hereafter, together with the strategies to further increase contrast in future devices { as the mirror edges and via make up a smaller fraction of the mirror total area}. 

a) Scattered light\\	
TI DMD design has a band of ``pond'' mirrors surrounding the active array of the DMD that are permanently landed in the OFF position.  These mirrors represent a potential source of scattered light, as they can reflect light into the spectroscopic or imaging channel especially when the instrument is blasted with light, which is normally the case { when} taking calibration exposures. These mirrors are masked close to the DMD surface but may receive light if e.g. the calibration optics do not perfectly match the f/\# and focal plane of the telescope.
Other sources of scattered light include the individual mirror edges and the “via”, the small central structure with thin walls supporting the mirror on top of the torsion hinge\cite{piotrowski_optical_2022}. Increasing the mirror size is expected to decrease the relevance of these factors.

b) Diffraction\\
When the size of the mirrors is comparable to the wavelength, diffraction becomes an issue. 
For astronomical applications coherence is spatially limited to the size of the PSF thus diffraction effects depend mostly on the “slit” used, i.e. the number of the illuminated mirrors in the "on" state. When a single mirror is used, some diffraction effect may still be caused by the adjacent “off” mirror edges and the underlying structure.  In general, the DMD efficiency for a single ON-state micromirror is maximized at short wavelengths and fast focal ratios \cite{piotrowski_optical_2023}. This drives to larger size the design of DMD mirrors optimized for astronomy. 			

\item{Packaging\\}
DMD packaging is designed  for projection applications where DMDs are blasted with extremely bright light with very fast f/\#. Key factors are the angle at which the incident light enters the package (side illumination is typical) and the positioning of the field stop surrounding the active area. Also the distance between the window and the mirror surface is a factor: a short distance enhances the window's defects (and therefore the cost to procure high purity glass), while a longer distance makes the device more prone to scattered light. Irrelevant for projectors but most important for astronomical applications, DMDs cannot be butted together, hence the complex optical design solutions devised for instruments requiring a very large number of spectra or wide field coverage\cite{content_offspring_2008,spano_dmd_2009}.

\item{Operations\\}
The fast pixel and frame rate typical of digital imaging applications drive complex electronic control circuitry.  The refresh rate of  high performance DMDs like the CINEMA  device requires hundreds of connecting lines with high-performance electronics placed in the immediate vicinity. This has negative  implications in terms of heat dissipation and reliability especially in extreme environments such as those encountered in space or, on the ground, in the IR. Building SAMOS, our JHU team has reversed engineered the TI DMD control board\cite{hope_digital_2018}. Future devices optimized for astronomy will have very relaxed dynamic requirements, allowing the electronic circuitry to be greatly simplified, with lower power consumption.
\end{enumerate}

\subsection* {Acknowledgments}
M.R. is indebted to Tim Heckman at JHU for encouraging to look into the possibility of creating ex-novo a future generation of micromirror devices specifically tailored for astrophysical research. 

\subsection* {Code and Data Availability Statement}
Data sharing is not applicable to this review article, as no new data were created or analyzed.

\subsection* {Disclosures}
The authors declare that there are no financial interests, commercial affiliations, or other potential conflicts of interest that could have influenced the objectivity of this research or the writing of this paper.


\bibliography{report.bib}   
\bibliographystyle{spiejour}   


\vspace{2ex}\noindent\textbf{M. Robberto} is Observatory Scientist at the Space Telescope Science Institute in Baltimore and Research Scientist at the Johns Hopkins University also in Baltimore. He is currently involved in the promotion and concept studies of the HWO, after having served for several years as Team Lead of JWST/NIRCam. 
Currently Dr. Robberto is also Principal Investigator of SCORPIO, the next facility instrument for Gemini South, and of SAMOS, an AO-fed multi-object optical spectrograph for the SOAR telescope. His research interests are mostly concentrated on the study of star formation.

\vspace{1ex}
\noindent Biographies and photographs of the other authors are not available.

\listoffigures
\listoftables

\end{document}